**Thiol-ene photo-click collagen-PEG hydrogels: impact of water-soluble photoinitiators on cell viability, gelation kinetics and rheological properties**

Róisín Holmes,[1] Xuebin B Yang,[1] Aishling Dunne,[2] Larisa Florea,[2] David Wood[1] and Giuseppe Tronci[1,3*]

[1] Department of Oral Biology, School of Dentistry, University of Leeds, Wellcome Trust Brenner Building, St James' University Hospital, Leeds, LS9, 7TF, UK

[2] Insight Centre for Data Analytics, National Centre for Sensor Research, School of Chemical Sciences, Dublin City University, Dublin 9, Glasnevin, Ireland

[3] Clothworkers' Centre for Textile Materials Innovation for Healthcare (CCTMIH), School of Design, University of Leeds, Leeds, LS2 9JT, UK

[*] Correspondence: g.tronci@leeds.ac.uk

**Abstract**

Thiol-ene photo-click hydrogels were prepared via step-growth polymerisation using thiol-functionalised type-I collagen and 8-arm poly(ethylene glycol) norbornene terminated (PEG-NB), as a potential injectable regenerative device. Type-I collagen was thiol-functionalised by a ring opening reaction with 2-iminothiolane (2IT), whereby up to 80 Abs% functionalisation and 90 RPN% triple helical preservation were recorded via 2,4,6-Trinitrobenzenesulfonic acid (TNBS) colorimetric assay and circular dichroism (CD). Type, i.e. 2-Hydroxy-1-[4-(2-hydroxyethoxy) phenyl]-2-methyl-1-propanone (I2959) or lithium phenyl-2,4,6-trimethylbenzoylphosphinate (LAP)] and concentration of photoinitiator were varied to ensure minimal photoinitiator-induced cytotoxicity and to enable thiol-ene network formation of collagen-PEG mixtures. The viability of G292 cells following 24 h culture in photoinitiator-supplemented media was largely affected by the photoinitiator concentration, with I2959-supplemented media observed to induce higher toxic response (0.1 → 0.5 % (w/v) I2959, cell survival: 62 → 2 Abs%) compared to LAP-supplemented media (cell survival: 86 → 8 Abs%). In line with the in vitro study, selected photoinitiator concentrations were used to prepare thiol-ene photo-click hydrogels. Gelation kinetics proved to be largely affected by the specific photoinitiator, with LAP-containing thiol-ene mixtures leading to significantly reduced complete gelation time ($\tau$: 187 s) with respect to I2959-containing mixtures ($\tau$: 1683 s). Other than the specific photoinitiator, the photoinitiator concentration was key to adjust hydrogel storage modulus (G'), whereby 15-fold G' increase (232 → 3360 Pa) was observed in samples prepared with 0.5% (w/v) compared to 0.1% (w/v) LAP. Further thiol-ene formulations with 0.5% (w/v) LAP and varied content of PEG-NB were tested to prepare photo-click hydrogels with porous architecture, as well as tunable storage modulus (G': 540 – 4810 Pa), gelation time ($\tau$: 73 – 300 s) and swelling ratio (SR: 1530 – 2840 wt.%). The photoinitiator-gelation-cytotoxicity relationships established in this study will be instrumental to the design of orthogonal collagen-based niches for regenerative medicine.

## 1. Introduction

The design of regenerative devices relies heavily on the use of three-dimensional (3D) scaffolds to provide the appropriate environment, mechanical support and an initial cell anchorage site for the regeneration of tissues and organs.[1] As a biomaterial, hydrogels provide a 3D hydrated framework with tissue-like elasticity for culturing cells. They also possess the ability to encapsulate cells and molecules prior to gelation, thus affording a minimally invasive avenue to deliver cells and bioactive factors.[2-4]

Collagen type I is a natural polymer and is the most abundant structural building block in connective tissues such as bone, tendon and cartilage.[5, 6]

The basic structure of collagen is a triple helix of three left-handed polypeptide chains, which can self-aggregate to form collagen fibrils, fibres and fascicles *in vivo.*[6-8] As a scaffold material, collagen is inherently biocompatible and bioactive, and has macromolecular organisation similar to the natural extracellular matrix (ECM), however its use as a biomaterial building block may be limited due to batch-to-batch variability, potential antigenicity and difficulty in tuning features such as stiffness, degradation and bioactivity.[9]

Cross-linking methods are necessary to stabilise collagen in aqueous solutions *ex vivo* and can be divided into three groups*:* physical interactions (e.g. ionically cross-linking and hydrogen bonds), chemical reactions (e.g. with glutaraldehyde,[10] diisocyanates,[11] carbodiimide-activated diacids,[12] or photochemical polymerisation[13]) or enzyme (e.g. glutaminase) catalysed mechanism.[14, 15]

Chemical cross-linking can often be performed via covalent modification of lysine and hydroxylysine residues and amino termini, cumulatively counting for ~ $3 \times 10^{-4}$ mol·g$^{-1}$ of collagen.[16-20] An advantage of chemical cross-linking is that the mechanical performance of the resulting hydrogel can be systematically adjusted via variation in cross-linker type and cross-link density, so that controlled tissue-specific ECM analogues can be successfully obtained.[9, 21, 22] Limitations include reagent toxicity, side reactions and lack of preservation of collagen triple helical conformation in resulting cross-linked systems, the latter being key to ensure high affinity for cells and mechanical stability in physiological conditions.[12, 23]

'Click' chemistry is a cross-linking method which invokes high reactivity and selectivity.[24-27] Other attributes of click chemistry include insensitivity to oxygen and water, high yields, mild reaction conditions and no generation of potentially harmful by-products, thus reducing the toxicity of the resulting product.[28, 29] As such, click chemistry is an appealing approach for the formation of collagen hydrogels.

Among click reactions, thiol-click chemistry has been shown to be a powerful tool for the efficient formation of thioether linkages (i.e., C–S–C).[30] Examples of two thiol-click reactions include: Michael-addition, which requires an electron-deficient carbon-carbon double bond (vinyl) and a base[31]; and the thiol-ene reaction, which proceeds via a radical mediated, step-growth mechanism between a thiol and a vinyl group that can be initiated using a light source and a photoinitiator.[32, 33] Here, the vinyl group should be incapable of homo-polymerisation to ensure the only reaction taking place is the thiol-ene click reaction, such being the case for vinyl ether or norbornene-based mono-/macromers.[34]

As a result of the reaction selectivity and oxygen insensitivity, an interesting feature of thiol-ene mixtures is their potential to be delivered as injectable liquid promptly leading to the formation of a tissue cavity-conformable hydrogel following application of a suitable external stimulus. In light of the quick gelation capability, thiol-ene mixtures can also be formulated with cells and molecules prior to gel formation, thus providing a potential avenue to deliver cells and bioactive factors in a minimally invasive manner.[35, 36] However, to achieve collagen gelation *in situ*, a major hurdle for the thiol-ene photo-click reaction is the necessary use of a non-toxic and water-soluble photoinitiator.[37, 38] Upon light exposure, the photoinitiator generates free radicals that may cause cellular damage at varied extent, depending on the gelation kinetics of the reacting mixture.[39] Additionally, most of the commercially-available photoinitiators are intended for use in organic solvents or dental resins and are therefore mostly insoluble in aqueous solutions of collagen.

Irgacure™ 2959 (I2959) and Lithium phenyl-2,4,6-trimethylbenzoylphosphinate (LAP) are two water-soluble photoinitiators capable of producing radicals when exposed to ultraviolet (UV) light (365 nm). I2959 is a commonly used photoinitiator for encapsulation of cells within hydrogels due to its well-established cytocompatibility at low concentrations.[40]

In this work, we explored the challenge associated with the formation of an injectable, regenerative device via a thiol-ene photo-click reaction in the presence of 2-iminothiolane (2IT)-functionalised collagen and 8-arm norbornene-terminated poly(ethylene glycol) (PEG-NB). 2IT was selected as a flexible thiol-forming compound and was covalently-coupled to type I rat tail collagen via ring-opening reaction using the amine group from (hydroxy-)lysine terminations. PEG-NB was employed as the synthetic hydrogel building block with the NB function providing the carbon-carbon double bond incapable of homo-polymerisation required for the thiol-ene reaction. The presence of NB-terminated multi-arms was expected to minimise reaction-hindering steric effects, thereby promoting faster gelation compared to linear PEG-NB variants. We hypothesised that the orthogonality offered by the thiol-ene reaction could result in the rapid formation of a defined collagen based hydrogel network (~ seconds) and therefore minimal toxic response.[41] Although thiol-ene hydrogels have been reported, none have been prepared using collagen as the precursor material.[27, 34, 36, 42, 43] The introduction of a PEG phase in the collagen-based hydrogel provided improved user control of gel formation and mechanical properties.[41, 44] Two water-soluble photoinitiators were employed to establish the photoinitiator-gelation-cytotoxicity relationship involved in the formation of thiol-ene photo-click collagen-PEG hydrogels.

## 2. Materials and Methods

### 2.1 Materials

2-iminothiolane hydrochloride (2IT), Dulbecco's modified eagle medium (DMEM) and penicillin streptomycin (Pen Strep) were purchased from Fischer Scientific Ltd. 2-Hydroxy-1-[4-(2-hydroxyethoxy)phenyl]-2-methylpropan-1-one (I2959) was purchased from Fluorochem Limited. 1,4-Dithiothreitol (DTT) and Foetal bovine serum (FBS) were purchased from Sigma. Lithium phenyl-2,4,6-trimethylbenzoylphosphinate (LAP) was purchased from Tokyo chemicals industry. 8-arm PEG Norbornene ($M_w$: 20,000 g mol$^{-1}$) was purchased from JenKem Technology USA. ATPlite luminescence assay was purchased from PerkinElmer.

### 2.2 Preparation of photo-click collagen-PEG hydrogels

#### 2.2.1 Extraction of collagen type I

Collagen type I was isolated in-house via acidic treatment of rat-tail tendons. Briefly, rat-tails were defrosted in ethanol before the skin was removed using a scalpel. The tails were left to dry before the exposed tendons (approx. four per tail) were removed, sliced and placed in acetic acid (17.4 mM) and stirred for 48 hours. The solution was centrifuged and the pellet removed to leave the soluble collagen type I dissolved in the acetic acid, which was hereafter freeze-dried to produce the white, acid-soluble collagen type I.

#### 2.2.2 Synthesis of collagen-SH

Collagen type I was stirred in acetic acid (17.4 mM, 1 wt.%) until dissolution. The pH of the solution was adjusted to pH 7.4 and then 2-iminothiolane (2IT) (5-20:1 [2IT]: [Lys]) was added with 1,4-dithiothreitol (DTT) (1:1 [DTT]: [2IT]). The solution was stirred for 24 hours before it was precipitated in ethanol (20-fold volume excess), centrifuged and the residue was air dried (Scheme 1). Functionalisation was confirmed using a (2,4,6) Trinitrobenzenesulfonic acid (TNBS) colorimetric assay.

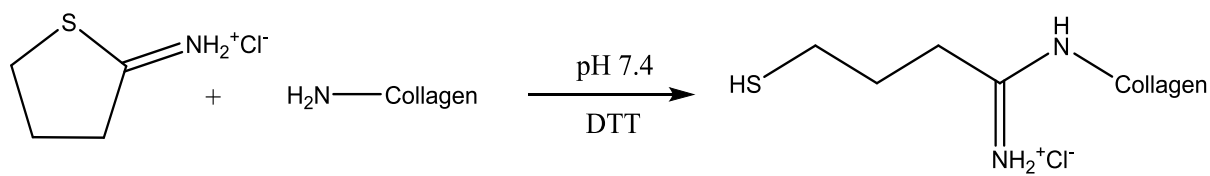

**Scheme 1.** Functionalisation of the available amine group in collagen with 2-iminothiolane (2IT).

### 2.2.3 Thiol-ene reaction

Collagen-SH (1 wt.%) was dissolved in a PBS solution containing either LAP or I2959 (0.1% (w/v) or 0.5% (w/v)). PEG-NB (2 -4 % (w/v)) was added to the collagen solution. Exposure to UV light (Spectroline, 365 nm) resulted in complete gelation by a thiol-ene step-growth reaction mechanism (scheme 2).

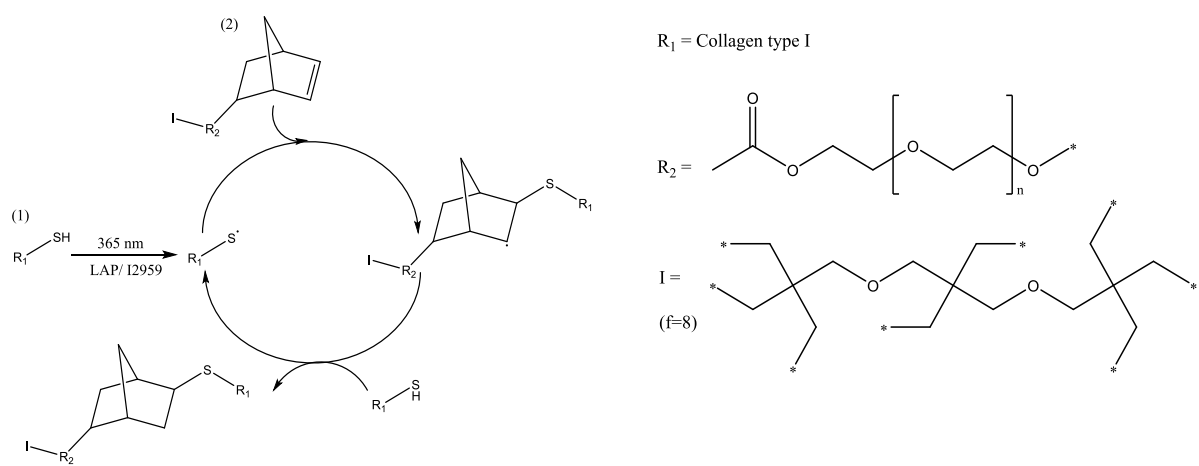

**Scheme 2.** Illustration of the thiol-ene photo-click reaction mechanism between collagen-2IT (1) and PEG-NB (2). Radical-mediated step-growth thiol-ene network formation was initiated via UV activation of aqueous solutions containing 2IT-functionalised collagen precursor and 8-arm (f=8) norbornene-terminated PEG in presence of either LAP or I2959 photoinitiator.

## 2.3 Chemical characterisation

### 2.3.1 2,4,6-Trinitrobenzenesulfonic acid (TNBS) colorimetric assay

Functionalised collagen (0.011 g) including a reference sample were placed in a vial. Sodium hydrogen carbonate ($NaHCO_3$) (4%, 1 mL) and TNBS (0.5%, 1 mL) were added. Hydrochloric acid (HCl) (6 N, 3 mL) was added to the reference sample. These were stirred at 40 °C for 3 hours. HCl (6 N, 3 mL) was added to the non-reference sample and stirred at 60 °C for 1 hour to complete the reaction. Dilution in water (5 mL) was carried out, followed by extraction in diethyl ether (3 x 15 mL). An aliquot (5 mL) was removed and diluted in water (15 mL). Absorbance was measured at 346 nm. The molar content of (hydroxyl-) lysine primary amino groups in either native or functionalised collagen was calculated using equation 1.

$$Mol\ (Lys) = \frac{2 \cdot Abs(346\ nm) \cdot 0.02L}{1.46\ x\ 10^4\ (M^{-1}cm^{-1}) \cdot b \cdot x} \tag{1}$$

Where *Abs(346 nm)* is the absorbance value at 346 nm, *0.02* is the volume of sample solution (in litres), *1.46 x 10$^4$* is the molar absorption coefficient for 2,4,6-trinitrophenyl lysine (in $M^{-1} \cdot cm^{-1}$), *b* is the cell path length (1 cm) and *x* is the sample weight. Accordingly, the degree of collagen functionalisation, *F*, was determined by TNBS colorimetric assay from the following equation:

$$F = 100 - \frac{Mol\ (Lys)_{2IT}}{Mol\ (Lys)_{coll}}\ x\ 100 \tag{2}$$

Where *Mol(Lys)$_{coll}$* and *Mol(Lys)$_{2IT}$* represent the total molar content of free amino groups in native (Lys ~ 3 · 10$^{-4}$ mol·g$^{-1}$) and 2IT-reacted collagen respectively.[16, 17]

### 2.3.2 Circular dichroism

Circular dichroism (CD) spectra of functionalised samples were acquired with a ChirascanCD spectrometer (Applied Photophysics Ltd). Solutions of functionalised collagen (0.2 mg.ml$^{-1}$) dissolved in acetic acid (17.4 mM) were collected in quartz cells of 1.0 mm path length. CD spectra were obtained using 4.3 nm band width and 20 nm min$^{-1}$ scanning speed. In addition, a spectrum of the acetic acid (17.4 mM) control solution was recorded and subtracted from each sample. Molar ellipticity ($\theta_{mrw,\lambda}$) was calculated from equation 3.

$$\theta_{mrw,\lambda} = \frac{MRW \cdot \theta_\lambda}{10 \cdot d \cdot c_d} \qquad (3)$$

Where $\theta_\lambda$ is the observed molar ellipticity (degrees) at wavelength $\lambda$, $d$ is path length (1 cm) and $c$ is the concentration (0.2 mg·ml$^{-1}$).[45]

The ratio of the magnitude of the positive to negative peaks (RPN) in the CD spectra was used as an indication of the triple helix architecture. For collagen functionalised samples, the quantification of the triple helix preservation ($X_c$) was calculated by normalisation of the RPN value with respect to the RPN value of native collagen (Literature value of collagen type I RPN 0.117).[46]

### 2.3.3 Ultra violet/ visible (UV/Vis) light spectroscopy

UV/Vis spectroscopy was carried out using both a Thermo Scientific NanoDrop Lite spectrophotometer and a JENWAY 6305 spectrophotometer. The solution concentration of either photoinitiators or collagen was adjusted to achieve absorbance values equal or under 1.

### 2.4 In vitro cytotoxicity assay

G292 cells (1 x 10$^4$ cells in 100 µL) were cultured in basal medium (Gibco$^{TM}$ DMEM, 10% FBS, 1% Pen Strep) in 96-well plates (n= 7). For the test groups, the cells were exposed to basal media with increasing concentrations of either I2959 or LAP (0.01% - 0.5% (w/v)) and cultured for 24 hours. Basal medium alone was used as the control and cell viability was assessed via an ATPlite assay. The absorbance of the experimental groups was normalised with respect to the corresponding basal medium control group to provide relative survival rates.[47, 48]

## 2.5 Rheology studies

Rheology curing measurements were performed on the thiol-ene reacting mixtures. The measurements were carried out on an Anton Paar MCR 301 rheometer using a CP50-2 tool with a diameter of 49.97 mm and a cone angle of 1.996°. UV light (365 nm) curing was initiated after 60 seconds with a light intensity of 4450 µW·cm$^{-2}$ at 25 °C. The time point at which the reaction was complete and a full chemical gel was formed was measured at the maximum recorded value of storage modulus (G'$_{max}$). The gelation time ($\tau$) was quantified as the temporal interval between UV activation and complete chemical gel formation. Amplitude sweep tests were performed on photo-cured hydrogels using the PP15 parallel plate tool at an angular frequency of 100 rad s$^{-1}$ and a normal force of 1 N.

## 2.6 Network architecture

### 2.6.1 Swelling ratio and gel content

Dehydrated samples (2 – 5 mg) were individually placed in 1 ml of distilled water at 25 °C under mild shaking. At specific time points, the swelling ratio (SR) was calculated using the following equation:

$$SR = \frac{M_s - M_d}{M_d} \; x \; 100 \qquad (4)$$

Where $M_s$ and $M_d$ are the swollen and dried weight. Swollen samples were paper blotted prior to weight measurement to consider the contribution from bound water only.

In addition to the swelling ratio, the gel content was determined to investigate the overall portion of the covalent hydrogel network insoluble in 10 mM HCl solution. Dried hydrogel networks of known mass ($M_d$) were equilibrated in 10 mM HCl solution for 24 hours. Resulting hydrogels were air dried and weighed ($M_{d1}$). The gel content (G) was calculated using equation 5.

$$G = \frac{M_{d1}}{M_d} \; x \; 100 \qquad (5)$$

**2.6.2 Morphology study**

Hydrated gels were investigated in a cool-stage chamber of a Hitachi S-3400N VP-SEM. SEM images were captured via backscattered electron detection at 5 kV and 12 – 13 mm working distance.

**2.7 Statistical analysis**

One-way ANOVA followed by post-hoc Tukey test on data when Levene's test showed p > 0.05 and equal variances could be assumed and a Welch's ANOVA followed by a post-hoc Games-Howell on data when Levene's test p < 0.05 and equal variances couldn't be assumed. Statistical significance was determined by p < 0.05.

**3. Results and Discussion**

The design of the thiol-ene photo-click collagen-PEG hydrogel system is presented, with the aim to identify key reaction parameters to enable fast gelation (~ seconds), minimal photo-initiator-induced toxic response and tunable hydrogel elasticity. Sample nomenclature used in this work is as follows: thiol-ene hydrogels are coded as "CollPEGX", where "Coll" and "PEG" refer to 2IT-functionalised collagen and 8-arm norbornene-terminated PEG, and "X" identifies the PEG content (2-4 % (w/v)) in the thiol-ene mixture.

**3.1 Synthesis and evaluation of collagen-SH**

Thiolated collagen was prepared via a base-free (hydroxyl-)lysine mediated ring-opening reaction using 2IT, which provided minimal impact on the charge properties of the original amino groups (Scheme 1). The occurrence of the reaction was confirmed via TNBS assay. This assay provides information on the molar content of the remaining free primary amino groups in reacted, compared to native proteins (Figure 1). The reaction could be easily and systematically controlled so that increasing the molar ratio of 2IT with respect to collagen primary amino functions ([2IT]: [Lys]) directly affected the degree of functionalisation in the resulting products (F: 55 → 80 Abs%).

The degree of triple helix preservation ($X_c$) of the collagen after functionalisation was determined using CD with normalisation against the RPN value of native collagen. The functionalised collagen displayed high preservation of triple helices ($X_c > 80$ RPN%)

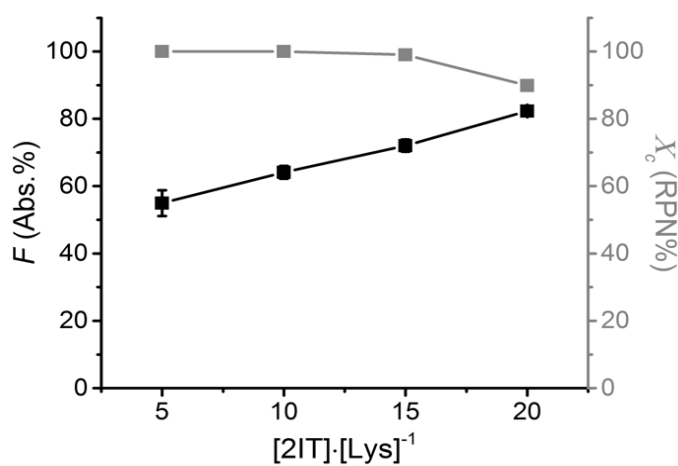

**Figure 1.** Degree of functionalisation (F) and triple helix preservation ($X_c$) in collagen products reacted with 2IT at varied [2IT]: [Lys]$^{-1}$ molar ratios. Data presented as mean ± SEM.

regardless of the degree of F(Abs.%), with almost quantitative yield ($X_c$: 90 RPN%) in 2IT-reacted products with a degree of functionalisation of up to 72 Abs% (Figure 1); this is beneficial to preserve the native protein organisation and high natural affinity for cells.[23] For further experiments, collagen-SH was prepared using 15 times excess (F: 72 Abs%, $X_c$: 90 RPN%)

**3.2 Photoinitiator analysis**

Following the formation of thiolated collagen, the effect of the water-soluble photoinitiators on cell viability was investigated. Minimal photoinitiator-induced cytotoxic response is key to enable radical-mediated thiol-ene gelation *in situ* in the presence of cells.

Photoinitiator toxicity was evaluated using an ATPlite assay on G292 cells cultured with an increased concentration of either LAP or I2959 (0.005 - 0.5% (w/v)) in basal medium for 24 hours. The relative survival rate of the cells was calculated by normalising the absorbance readings of the samples against the control, which was G292 cells grown with basal medium on cell culture plastic using seven repeats (Figure 2.A). The results showed that there was no detected cell toxicity after culturing the cells for 24 hours in cell culture media with photoinitiator concentrations below 0.05% (w/v) for both LAP and I2959. However, the relative survival was observed to significantly decrease (p <0.05) when the photoinitiator concentration in cell culture media was increased from 0.1 →0.5 % (w/v) for both LAP (cell survival: 86 → 8 Abs%) and I2959 (cell survival: 62 → 2 Abs%). At the same time, the results showed a significant difference in cell viability depending on the specific photoinitiator supplemented in the basal media, with LAP outperforming I2959 (p<0.05).

In addition to the impact on cell viability, the radical-generation capability of both photoinitiators using UV light (365 nm) was evaluated. The UV-Vis spectra of I2959 and LAP photoinitiators and collagen are presented (Figure 2.B). The maximum absorption wavelength of LAP was found to be 380 nm, with a high absorbance still recorded at 365 nm (molar extinction coefficient ($\varepsilon$): 218 $M^{-1}$ $cm^{-1}$).[47] In comparison, I2959 had a peak maximum at 285 nm, where it also displayed competing absorbance against the peak from the amino acids in collagen, whilst displaying minimal absorbance at 365 nm ($\varepsilon$: 4 M$^{-1}$ cm$^{-1}$).[47]

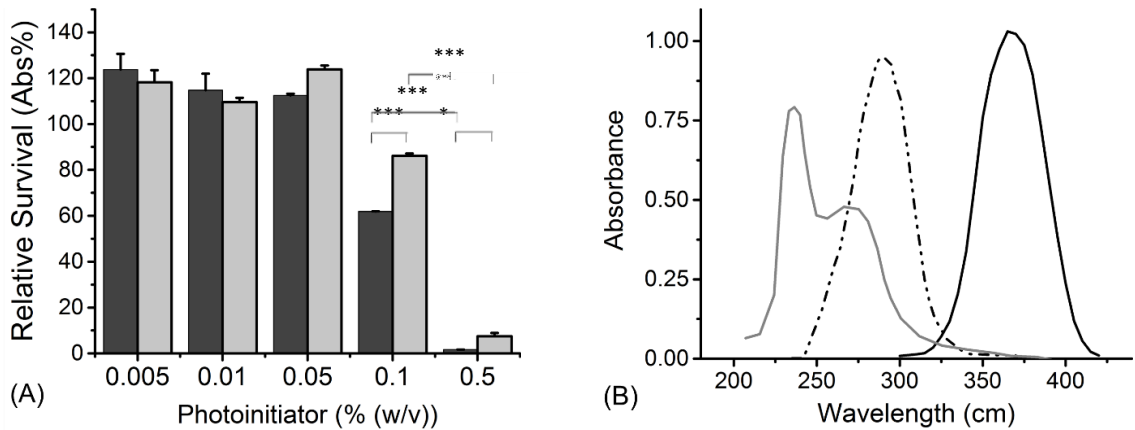

**Figure 2.** Initiator analysis. (A): Relative survival of G292 cells cultured in cell culture medium supplemented with increasing initiator concentrations. I2959 (●); LAP (●). Cell survival was quantified using an ATPlite viability assay. Data are presented as means ± SEM. * P<0.05; ***P<0.001. (B): UV/Vis spectrum showing the absorbance of collagen, LAP and I2959 solutions. Collagen (—); I2959: (---); LAP (—).

Photopolymerisation reactions are driven by chemicals that generate free radicals when exposed to specific wavelengths of light. The rate of a photopolymerisation reaction is directly proportional to the photoinitiator concentration, light intensity, the molar extinction coefficient and photoinitiator efficiency.[49] When carried out in an aqueous environment, these reactions commonly use photoinitiator concentration above 0.1% (w/v) to ensure efficient generation of radicals for the reaction to occur; although it has been reported that when incorporated with peptides, I2959 does not initiate photopolymerisation reactions as effectively at concentrations below 0.3 % (w/v).[48, 50-52]

The weak absorbance profile of I2959 at 365 nm and its competing absorbance with collagen severely limits its utility for photopolymerisation reactions performed at this wavelength. This implies that a higher concentration of I2959 is needed when used at 365 nm compared with LAP to overcome these issues. On the other hand, the low solubility of I2959 in water together with considerations on respective photoinitiator-induced cytotoxic effect make this approach unfeasible. Considering previously observed trends of solubility, absorbance profiles and cell viability, photoinitiator concentrations in the range of 0.1-0.5 % w/v were employed to induce radical-mediated gelation of thiol-ene collagen-PEG mixtures.

### 3.3 Thiol-ene gelation kinetics via rheological studies

Confirmation of thiol-ene hydrogel formation was obtained via rheology curing time sweep measurements during which storage (G') and loss (G'') moduli were recorded. The gelation kinetics profile of thiol-ene mixture, CollPEG4 (LAP/ 0.5% (w/v) is displayed in Figure 3. Comparable to the liquid state, the collagen-PEG mixture was initially found to be predominantly viscous.

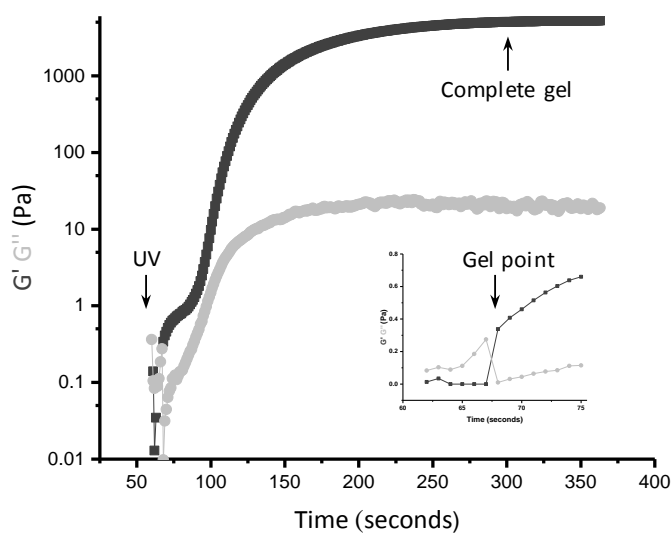

**Figure 3.** Typical gelation kinetics profile of thiol-ene mixture CollPEG4 (LAP 0.5% (w/v)). UV light was activated 60 seconds after the start of the time sweep measurement, resulting in complete gelation after nearly 300 seconds. The inner image presents a zoomed-in plot of the initial stage of gel formation, indicating a gel point after just 7 seconds of UV activation.

The UV light was turned on 60 seconds after the start of the time sweep, which promptly induced the formation of a covalent network, and is clearly indicated by the detection of a gel point (the time at which the storage and loss moduli equate or crossover) seven seconds after the photo-activation.[53, 54] The storage modulus was found to increase with time and UV exposure until a plateau was reached within 300 s, indicating no further elastic properties (complete chemical gel). The gelation kinetics of the collagen-PEG thiol-ene mixture prepared with either LAP or I2959 with varied photoinitiator concentration was also characterised. The results from Table 1 show a significant difference in gelation time depending on choice of photoinitiator introduced in the thiol-ene solution. I2959-based hydrogel-forming solutions took approximately 8 times as long to reach gel completion compared to LAP-containing thiol-ene solutions, regardless of photoinitiator concentration. The slower gelation rate displayed by hydrogel-forming thiol-ene mixtures prepared with I2959 compared to LAP could be due to competing absorbance at 365 nm with collagen and the specific wavelength used, as discussed previously. Other than the gelation time, the choice of photoinitiator had no comparable effect on the storage modulus (Table 1). At 0.5% (w/v), LAP and I2959 presented an average storage modulus of 3360 Pa and 3029 Pa; whilst at 0.1% (w/v) photoinitiator, LAP and I2959 presented a storage modulus of 232 Pa and 190 Pa. Despite the photoinitiator-induced variation in gelation kinetics, this data suggests comparable cross-link density among previously-mentioned hydrogel formulations, as expected, considering the selectivity and oxygen-insensitivity of the thiol-ene click reaction.

**Table 1.** Effect of photoinitiator type and concentration on both gelation kinetics and storage modulus obtained of thiol-ene mixture CollPEG3.5 as investigated via time and amplitude sweeps, respectively. Data presented as means ± SEM.

| Photoinitiator (% w/v) | G′ (Pa) | $\tau$ (s) |
|---|---|---|
| I2959 (0.1) | 190 ± 22 | 1496 ± 43 |
| I2959 (0.5) | 3029 ± 100 | 1683 ± 33 |
| LAP (0.1) | 232 ± 39 | 279 ± 11 |
| LAP (0.5) | 3360 ± 91 | 187 ± 6 |

Table 1 showed that increasing the photoinitiator concentration from 0.1% to 0.5% (w/v) in the thiol-ene mixture resulted in a 15-fold increase in the storage modulus, for both photoinitiators. Compared to 0.5% (w/v), the photoinitiator efficiency at 0.1% (w/v) appears to be too low to efficiently release enough radicals to complete the click chemistry reaction, thus resulting in fewer thiol-norbornene cross-links and a hydrogel with reduced storage modulus.

Due to the enhanced gelation kinetics and hydrogel storage modulus, the remaining rheological studies were carried out on thiol-ene solutions containing LAP 0.5% (w/v), aiming to induce further control on gelation kinetics, as well as hydrogel mechanical and physical properties.

**Table 2.** Mechanical and physical properties of four click chemistry gels using LAP (0.5%) Data presented as means ± SEM.

| Sampled ID | G' (Pa) | G'' (Pa) | $\tau$ (s) | Tan δ ($\times 10^{-3}$) |
|---|---|---|---|---|
| CollPEG2 | 540 ± 23 | 5.1 ± 1.0 | 73 ± 3 | 5.3 ± 0.2 |
| CollPEG2.5 | 1150 ± 46 | 5.8 ± 0.1 | 110 ± 6 | 6.3 ± 0.1 |
| CollPEG3 | 2040 ± 50 | 8.6 ± 0.3 | 133 ± 6 | 5.7 ± 0.1 |
| CollPEG3.5 | 3360 ± 91 | 11.1 ± 0.6 | 187 ± 6 | 5.0 ± 0.1 |
| CollPEG4 | 4810 ± 41 | 13.2 ± 0.8 | 301 ± 13 | 3.5 ± 0.1 |

Table 2 shows that five thiol-ene photo-click samples displayed an increased storage and loss modulus with the increased concentration of PEG in the thiol-ene solution (p <0.05).The photo-click gels displayed a G' in the range of 500 – 4800 Pa depending on the thiol-ene formulation, thus indicating a significant degree of tunability in the mechanical properties of the gels. Tan δ is a representation of the ratio of the viscous (loss) and elastic (storage) moduli (G''/G'). This parameter identified how tacky/sticky the hydrogel was.[55] Interestingly, Tan δ was observed to decrease with increasing

PEG content, ultimately leading to increased hydrogel stiffness and reduced tackiness of the resulting gel. This data provides further evidence that PEG is integrated in a covalent network rather than acting as secondary plasticising phase, as also supported by respective gelation kinetics profiles (Figure 4).

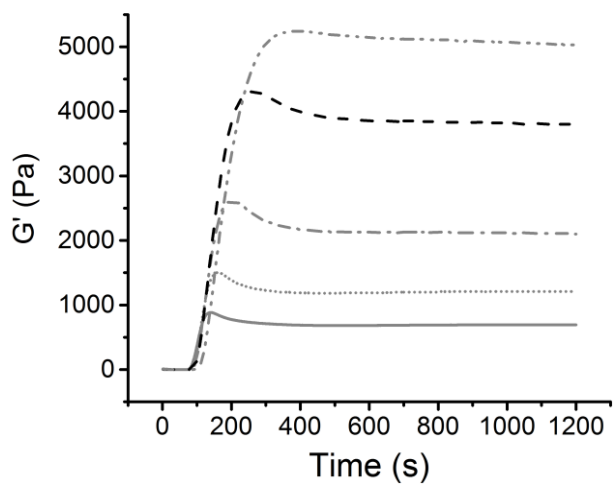

**Figure 4.** Curing time sweep measurements of thiol-ene collagen mixtures prepared with 0.5% (w/v) LAP and varied PEG content. CollPEG2: (—); CollPEG2.5: (⋯); CollPEG3: (-⋯-); CollPEG3.5: (- -); CollPEG4: (-⋯-).

### 3.4 Network Architecture

Swelling and gel content were quantified to gain insight into the network architecture of the photo-click collagen-PEG hydrogels. Other than the swelling ratio providing information on the cross-link density, the gel content was used to assess the portion of extractable material not involved in the formation of the thiol-ene photo-click covalent network.[21] As observed in Table 3, high gel content (G > 90 wt.%) was found for all the thiol-ene photo-click gels investigated, suggesting all the material was held together as a covalent network. This observation provided further confirmation that the presented synthetic thiol-ene approach successfully enabled the rapid formation of defined covalently-crosslinked collagen-based networks.

**Table 3.** Swelling ratio and gel content of the click chemistry gels. Data represented as mean ± SEM.

| Sample ID | SR (wt.%) | G (wt.%) |
|---|---|---|
| CollPEG2 | 1530 ± 130 | 90 ± 1 |
| CollPEG2.5 | 2130 ± 150 | 86 ± 3 |
| CollPEG3 | 2340 ± 240 | 97 ± 1 |
| CollPEG3.5 | 2640 ± 220 | 90 ± 2 |
| CollPEG4 | 2840 ± 71 | 91 ± 1 |

A paper by Xu et al reported the preparation of collagen hydrogels using a Michael-addition reaction.[56] These hydrogels were recorded to have a gel content of 77 wt.%, which is lower than the one measured on the thiol-ene hydrogels obtained in this study. This results suggest a higher yield of network formation compared to previously-reported Michael-addition hydrogels, although an 8-arm rather than linear norbornene-derived PEG was employed in this study.[57]

Other than the gel content, the swelling ratio increased from 1530 wt.% to 2840 wt.% from collPEG2 to collPEG4, despite the increased storage modulus observed for these samples. This increase in swelling ratio could be due to the increased content of hydrophilic PEG incorporated into the hydrogel network. In comparison, collagen hydrogels formed using N-(3-Dimethylaminopropyl)-N′-ethylcarbodiimide hydrochloride (EDC) (20 x excess) have had a reported swelling ratio of 1595%.[12] Swelling ratio is an important feature for the design of regenerative devices since a high swelling ratio is expected to promote increased diffusion of nutrients and cellular waste into and out of the collagen hydrogel.[58] To further analyse the swelling behaviour of the photo-click gels, a swelling kinetic experiment was used to assess the amount of time needed for the gels to reach equilibrium with water (Figure 5). The stiffer gels, CollPEG4 and CollPEG3.5, were found to reach equilibrium after 10 minutes' incubation in distilled water, whereas the weakest gel, CollPEG2, needed 360 minutes to reach equilibrium with water. These results further confirm the relationship between PEG content in the thiol-ene reacting mixture, and previously-observed swelling ratio of the resulting hydrogel.

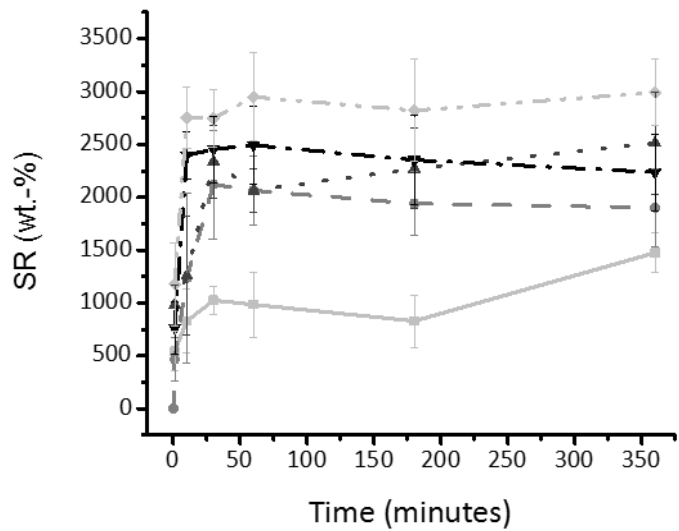

**Figure 5.** Swelling kinetics profile of thiol-ene collagen hydrogels prepared with 0.5% (w/v) LAP and varied PEG content. CollPEG2: (—); CollPEG2.5: (···); CollPEG3: (---); CollPEG3.5: (- -); CollPEG4: (-·-). Data presented as means ± SEM.

SEM images were taken in the wet state to analyse the morphology of the photo-click gels in near-physiological conditions (Figure 6). Pore sizes in the range of 30 – 80 μm were observed. Pores may likely be formed in the hydrogels following stirring and casting of the thiol-ene mixture. Pore size is an important feature in regenerative devices to allow sufficient material surface for cell seeding and cell diffusion. Additionally, pores should be large enough to promote cell diffusion. Additionally, pores should be large enough to promote vascularisation.[59] Work by Chiu et al (2011) stated that when using PEG hydrogels, cell and vessel invasion was limited to the external surface of the gels with smaller pore size (25 -50 μm), whereas gels with larger pores (50 – 150 μm) permitted mature vascularised tissue formation throughout the entire material volume.[60] The pore size measured in the photo-click hydrogels may therefore be suitable to enable angiogenesis and wound healing *in vivo*.

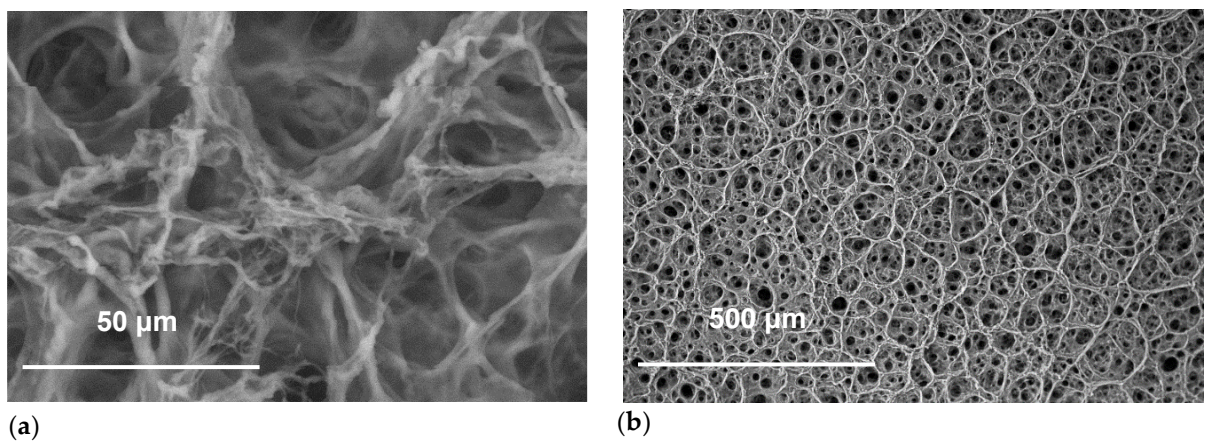

(a)　　　　　　　　　　　　　　　(b)

**Figure 6.** Cool-stage SEM images of sample CollPEG4. Magnifications: A: 1000×; B: 100x.

## 4. Conclusions

Collagen was successfully reacted with 2IT without the need for an additional base yielding thiolated collagen with a triple helical preservation of at least 90 RPN% (CD) and degree of functionalisation up to above of 80 Abs%. A thiol-ene click reaction was performed with PEG-NB in the presence of a water-soluble photoinitiator to investigate its potential as a rapid, oxygen-insensitive cross-linking method for *in situ* gelation. Photoinitiator-induced cytotoxicity as well as hydrogel gelation kinetics and storage modulus were examined using two water soluble photoinitiators, i.e. LAP and I2959, at varied photoinitiator concentrations (0.1 - 0.5% (w/v)). Photo-curing rheological experiments showed that decreased photoinitiator concentrations in the thiol-ene mixtures had a detrimental (15 fold) effect on storage modulus, whilst the lower, less toxic photoinitiator concentration was not feasible for gelation. I2959 photoinitiator was shown to be the least preferable photoinitiator when used at 365 nm, leading to longer gelation times compared to LAP.

Further work into using LAP 0.5% (w/v) showed that photo-click collagen-PEG hydrogels were successfully prepared with tunable physical and mechanical properties by altering the PEG content and thereby the number of thiol-ene crosslinks. Such soft and highly swollen systems offer widespread application in the context of e.g. muscle regeneration and wound healing devices.

**Acknowledgments**

This research was funded by the EPSRC Doctoral Training Centre in Tissue Engineering and Regenerative Medicine, a collaboration between the Universities of Leeds, Sheffield and York. Grant number EP/5000513/1. The authors wish to thank Professor Timothy F. Scott (University of Michigan, USA) for the scientific discussion at the beginning of the project. GT gratefully acknowledges financial support from the Wellcome Trust-University of Leeds Institutional Strategic Support Fund (ISSF) and the Clothworkers' Centre for Textile Materials Innovation for Healthcare (CCTMIH).

**Conflicts of Interest:** The authors declare no conflict of interest.